\documentclass[twocolumn]{aastex631}

\usepackage[T1]{fontenc}
\usepackage[utf8]{inputenc}
\usepackage{float}
\newcommand{\bpic}[1]{$\beta$ Pic}
\newcommand{\apic}[1]{$\alpha$ Pic}

\usepackage{graphicx}
\hypersetup{linkcolor=blue, citecolor=blue}
\usepackage{amsmath}

\shorttitle{Fomalhaut Planetesimal Population}
\shortauthors{Avsar et al.}

\begin{document}


\title{Forecasting Catastrophe: Constraints on the Fomalhaut Main Belt Planetesimal Population from Observed Collisional Remnants}

\author[0000-0001-7801-7425]{Arin M. Avsar}
\affiliation{Lunar and Planetary Laboratory, The University of Arizona, USA}
\author[0000-0002-4309-6343]{Kevin Wagner}
\affiliation{Steward Observatory, The University of Arizona, USA}
\author[0000-0003-3714-5855]{Dániel Apai}
\affiliation{Steward Observatory, The University of Arizona, USA}
\affiliation{Lunar and Planetary Laboratory, The University of Arizona, USA}
\affiliation{James C. Wyant College of Optical Sciences, The University of Arizona, USA}

\author{Christopher C. Stark}
\affiliation{NASA Goddard Space Flight Center, Exoplanets and Stellar Astrophysics Laboratory, Code 667, Greenbelt, MD 20771, USA}

\author[0000-0002-4388-6417]{Isabel Rebollido}
\affiliation{European Space Agency (ESA), European Space Astronomy Centre (ESAC), Camino Bajo del Castillo s/n, 28692 Villanueva de la
Canada, Madrid, Spain}


\begin{abstract}

Catastrophic planetesimal disruptions offer a unique opportunity to study and characterize large planetesimal populations in exoplanetary systems that are not currently detectable by modern observatories. The unexpected discovery of a second collision event in the Fomalhaut system raises important questions about the planetesimal population and dynamical state inside the Fomalhaut main belt that led to two collisions in 20 years. We present a statistical model developed and applied to the archetypal Fomalhaut system to provide new constraints on the bulk properties of the planetesimals in Fomalhaut’s main belt. Utilizing the constraints provided by the spatially resolved Fomalhaut cs1 and cs2 collision events, we retrieve the belt parameters that best reproduce the observed collision rate while remaining consistent with the system's age and dust mass. Our best-fit model suggests a total main belt mass of 200–360 $M_{\oplus}$, with the transition from a collisionally evolved to a primordial planetesimal population occurring at a radius of $115_{-10}^{+30}$ km and a maximum planetesimal radius of $380^{+643}_{-202}$ km. We estimate a catastrophic collision rate of $0.086_{-0.048}^{+0.067}$ collision events per year for planetesimals with radii $\ge$ 100 km in the region interior to the main belt. Our findings show that further observable collisions are likely, motivating continued monitoring of Fomalhaut and other nearby debris disks.
\end{abstract}


\section{Introduction}

Debris disks represent the rocky and icy remnants of planet formation, analogous to the asteroid and Kuiper Belts in our own Solar System. Micron$-$ to millimeter-sized dust grains have been imaged around at least 40 stars of varying ages and spectral types \citep[e.g.,][]{Engler2025}, and the observed dust is continually replenished through the collisional grinding of larger planetesimals \citep[e.g.,][]{Harper1984}. However, the properties of these parent bodies remain largely unconstrained as they lie below current detection limits \citep{Krivov2010,Hughes2018}. Previous modeling of these populations has relied on extrapolating size distributions from power-law indices governing the observed grains \citep[e.g.,][]{Holland2017}. However, this "bottom-up" extrapolation often yields total debris disk masses that exceeds the theoretical solid-mass inventory of the initial protoplanetary disk \citep{Krivov&Wyatt2021}.

To date, most candidate collision events have been inferred from unresolved infrared brightness variations in the inner few au of systems \citep[e.g.,][]{Su2019,Chen2024}. Only two planetary systems, $\beta$ Pictoris \citep{Dent2014,Apai2015,Rebollido2024} and Fomalhaut \citep{Kalas2008,Gaspar2020,Kalas2025}, have yielded spatially resolved collision remnants. This study focuses on Fomalhaut, where the collision remnants have been observed to have localized, time-varying surface brightness changes consistent with the aftermath of recent collisions between planetesimals.

Fomalhaut is an A3V main sequence star located at a distance of $7.70 \pm 0.03$ pc with a mass of $1.92 \pm 0.02 M_{\odot}$ \citep{vanLeeuwen2007,Mamajek2012}. It hosts an eccentric debris belt with a forced eccentricity of approximately 0.1 \citep[e.g.,][]{MacGregor2017}, in addition to a newly discovered intermediate belt and inner disk \citep{Gaspar2023}. Due to its proximity, Fomalhaut is one of the most comprehensively studied debris disks, imaged from optical to mm wavelengths \citep{Holland1998,Stapelfeldt2004,Kalas2005,Lebreton2013,Su2013,MacGregor2017,Adams2018,Chittidi2025}, and has long been a primary target for exoplanet searches \citep{Roman1959,Quillen2006,Chiang2009,Boley2012,Ygouf2024}.

There are no current confirmed exoplanets in the system. The first exoplanet candidate in the system, Fomalhaut b, was discovered in Hubble Space Telescope (HST) coronagraphic observations by \cite{Kalas2008}. However, its non-detection at infrared wavelengths \citep{Janson2012}, its potentially belt-crossing orbit \citep{Beust2014}, and the expansion and dimming of the candidate over time \citep{Gaspar2020} led to its reclassification as an expanding dust cloud created by a collision between unseen planetesimals.

New HST imaging in 2023 revealed a second collision remnant, Fomalhaut cs2, at the inner edge of the Fomalhaut main belt \citep{Kalas2025}. The discovery of a second catastrophic collision within a span of only 20 years was unexpected. Large planetesimal collisions in mature systems like Fomalhaut are theoretically rare, partially due to the extrapolated power-law used to model the planetesimal population, with expected occurrence rates on the order of one collision every $10^5$ years \citep[e.g.,][]{Wyatt2008}. The detection of two such events near the main belt in just two decades suggests such collision events are not anomalous. This would imply that a more complex planetesimal power-law distribution should be considered, and the planetesimal population is possibly more massive and dynamically active. This suggests that the Fomalhaut cs1 and cs2 events were not anomalous, but rather part of dynamical processes that can be used to probe the underlying planetesimal population of the belt.

In this study, we leverage these observed collisions as a proxy to constrain the total  mass and planetesimal population within the Fomalhaut main belt. We present a statistical model that calculates catastrophic collision rates for a given set of belt parameters, utilizing a ``top-down" approach enabled by the Fomalhaut cs1 and cs2 events. This approach offers an alternative to the standard "bottom-up" methodology \citep[e.g.,][]{Holland2017}, which relies on extrapolating the size distribution of large planetesimals from small dust grains. We use this model to retrieve the belt parameters that best reproduce the observed collisions in the Fomalhaut system, while remaining consistent with the system's age and dust mass inferred from ALMA observations. Furthermore, we use these best-fit parameters to predict the time and location of future observable collisions. Finally, we discuss how these findings can guide observations with future space-based coronagraphic missions, specifically in distinguishing between genuine Earth-like exoplanets and the false-positive signals generated by planetesimal collisions.

\section{Planetesimal Collision Model and Methodology}\label{sec:methods}

In this section, we will outline the assumptions and methodology we use in order to create the one-dimensional, axisymmetric planetesimal collision model that will be used to model the Fomalhaut main belt. We start by outlining important assumptions in the model, like the belt geometry, the density of the planetesimals, and the criteria we impose for a planetesimal to become disrupted. We then define the planetesimal size distribution the model employs and how a belt-wide collision rate is calculated. Finally, we outline how this model will be combined with the collision events and other observables in the Fomalhaut system to fit the underlying planetesimal population properties in the Fomalhaut belt.

 \subsection{Model Assumptions and Simplifications}

\subsubsection{Planetesimal Belt Geometry and Mass Distribution}\label{volume}

We begin with a planetesimal belt with an inner edge at $a_{min}$ and an outer edge at $a_{max}$. Furthermore, we define the peak location of the belt semi-major axis as $a_{belt}$, and the width of the belt as $\Delta a$.

We use an approximation for the volume of the planetesimal belt by assuming a low eccentricity (e $<< 1$) toroidal debris ring:

\begin{equation}\label{volume-eq}
    V \approx  2\pi a^2h\Delta a
\end{equation}

where $h$ is the vertical aspect ratio of the belt.

Additionally, we assume the planetesimals in the Fomalhaut belt follow the same spatial distribution of the dust observed with ALMA, resulting in a Gaussian surface density distribution \citep[e.g.][]{Kennedy2020} for planetesimals in the belt. In this case, the surface density peaks at $a_{belt}$, the location of the observed main belt as seen with ALMA \citep{MacGregor2017,Lovell2025} and falls off towards the edges of the disk boundaries, which are given by:

\begin{equation}
    \Sigma(r) \propto e^{- \frac{(r-a_{belt})^2}{2\sigma_r^2}}
\end{equation}

where $\sigma_r$ is proportional to the FWHM width ($\Delta a_{FWHM}$) of the disk as observed with ALMA.

\subsubsection{Assumptions about Planetesimal Dynamics}

Within the bounds of the debris belt, we assume an average relative velocity in which planetesimals encounter one another within the belt. This can also be defined as the free eccentricity of the planetesimals within the belt, where the free eccentricity is the deviation of the planetesimals from the eccentricity governing the overall planetesimal belt \citep[e.g.][]{MacGregor2017}. Since the eccentricity distribution is not well known for large planetesimals \citep[e.g.][]{Krivov2010,Hughes2018} in debris belts, and the relative velocity of the planetesimals is what governs the collision rate in the planetesimal belt (see Equation \ref{totalrate}), we choose to define an average relative velocity that will govern the collision rate instead of a free eccentricity distribution. We do not include specific orbital resonances associated with a hypothetical interior planet.

We note that this does not dictate that a single relative velocity governs the entire belt under this model. There can be sub-regions within the belt with differing average relative velocities that can indicate dynamically hot and dynamically cold regions of a planetesimal belt, as we will show in Section \ref{fomalhaut-config}.

\subsubsection{Radius-Density Relationship}\label{density}

We use an approximation from Kuiper Belt Objects (KBOs) with known densities. All grains and planetesimals in the model will follow the radius-bulk density relationship from \cite{Pearce2025} of:

\begin{equation}\label{eq2}
\frac{\rho(s)}{\text{g } \text{cm}^{-3}} \;=\;
\begin{cases}
0.8,
&  s \leq 100 \text{ km},
\\[1em]
0.08 \sqrt{\frac{s}{\text{km}}},
&  s > 100 \text{ km},
\\[1em]
\end{cases}
\end{equation}

where $s$ is the dust and planetesimal radii. This model provides a size dependent density for the planetesimals in the model that has been fit to known objects in the Solar System. This piecewise radius-dependent approach takes into account the low-density, icy nature of planetesimals in the outer Solar System, while also accounting for the compacting of large bodies under their own self-gravity \citep[e.g.][]{Bierson2019}. An accurate density model is important for collision models, as it determines the minimum impactor size needed to disrupt a target planetesimal (see Equation \ref{eq6}) and, in turn, the overall disruption rate of the target planetesimals.

\subsubsection{Planetesimal Disruption Criteria}\label{catastrophic-disruption}

We assume that the disrupted planetesimals that have been observed are catastrophic collisions in which the target planetesimal with radius $s_{targ}$ is broken apart by an impacting planetesimal with radius $s_{imp}$. We calculate $s_{\mathrm{imp}}$ for a target of ra $s_{targ}$ by equating the kinetic energy of the impactor to the binding energy of the target:

\begin{equation}\label{eq6}
    s_{\mathrm{imp}}(s) = [\frac{8 \pi}{5}\frac{G \rho_{\mathrm{targ}}^2 s_{\mathrm{targ}}^5}{v_{\mathrm{imp}}^2\rho_{\mathrm{imp}}}]^{1/3}
\end{equation}

where $v_{imp}$ is the impact velocity between the two objects, $\rho_{\mathrm{targ}}$ is the density of the target planetesimal, and $\rho_{\mathrm{imp}}$ is the density of the impacting planetesimal. Furthermore, we set $v_{imp} \approx v_{rel} $ \citep[e.g.][]{Armitage2010}.  Here we assume that the cohesion of bodies with $s > 0.1$ km is dominated by the object's self gravity and that material strength can be ignored in the size range of interest when calculating $s_{imp}$.

\subsection{Grain and Planetesimal Size-Distribution}\label{dist}

We use a dust and planetesimal population that contains three planetesimal populations distributed between radii of $s_{min}$ and $s_{max}$ with $2\times 10^4$ logarithmically-spaced radius bins following \cite{Rebollido2024}. The first two power-law indices follow that of a collisionally-evolved planetesimal population that is in steady state \citep{Lohne2008,Wyatt2011} where larger planetesimal are continuously being ground down into smaller grains and planetesimals. The third power law population encompasses the primordial population of planetesimals that have not yet been collisionally evolved.

In this scenario, from $s_{min}$ until 0.1 km in radius, we assume that the dust grains will follow the Dohnanyi size-distribution ($\alpha_1 = -3.5$), where material strength plays the primary role in the cohesion of small grains compared to gravity \citep{Dohnanyi1969}. At 0.1 km, the planetesimal population transitions ($\alpha_2 = -3$) from a collisional cascade from bodies held together by material strength to bodies mainly held together by their own self-gravity \citep{Pan2005,Lohne2008}. The planetesimal population follows this power law until it reaches $s_{break}$, where it then transitions from a collisionally evolved population to the primordial population. Finally, from $s_{break}$ to $s_{max}$ the power-law index transitions to $\alpha_3$. In this regime, the population is made up of large planetesimals and dwarf planets representing the primordial population of planetesimals that have not been collisionally disrupted.

The final size-distribution is as follows:

\begin{equation}
N(s) \;\propto\;
\begin{cases}
s^{\alpha_1},
& s_{min} < s \leq 0.1 \text{ km},
\\[1em]
s^{\alpha_2},
&  0.1 \text{ km} < s \leq s_{\mathrm{break}},
\\[1em]
J \times s^{\alpha_3},
& s_{\mathrm{break}} < s \leq s_{max},
\\[1em]
\end{cases}
\end{equation}

The primordial depletion parameter ($J$), is a unitless parameter that dictates the offset of the discontinuity between the collisionally evolved population and the primordial population. \cite{Wyatt2011} argues that the planetesimal size distribution transition between the collisionally evolved and primordial population does not have to be continuous and may be discontinuous for a collisionally evolved population. A $J$ of unity would imply that a continuous transition from the collisionally evolved to the primordial planetesimal population, while $J \neq 1$ implies a discontinuous transition.

We choose a $s_{min}$ = 1 $\mu$m, and, we find that changing the choice $s_{min}$ has a negligible impact on collision rates of large planetesimals. More specifically, we find that changing the choice of $s_{min}$ by an order of magnitude in either direction changes the collision rate of planetesimals by $< 0.1 \%$. However, the same cannot be said about $\alpha_3$, $s_{max}$, and $s_{break}$. In the coming sections, we will explore how varying each parameter impacts collision rates between planetesimals.

\subsection{Normalizing Distributions to the Total Disk Mass}

Once a planetesimal population is generated from a given size-distribution, we then normalize the population to a given belt mass. This is a key step in the model as the belt mass will play a critical role in determining the number density and collision rate of large planetesimals (see Equation \ref{single-rate}). This requires knowledge of the belt mass ($M_{belt}$), total volume of the planetesimal belt ($V$) as defined in Section \ref{volume}, and density distribution of the objects in the belt ($\rho$) as defined in Section \ref{density}.

In order to scale the number of objects in each bin to sum to the total disk mass, we generate a population of planetesimals with an arbitrary mass scaling, $N_{initial}(s)$, that has the intended shape of the planetesimal distribution. We then normalize to a given belt mass using the following expression:

\begin{equation}
    N(s) = N_{initial}(s) \times \frac{M_{belt}}{\sum_{s=s_{min}}^{s_{max}}\frac{4}{3}\pi \rho(s)s^3 N_{initial}(s)}
\end{equation}

\subsection{Planetesimal Collision Model}\label{model}

We calculate the rate at which an object with radius $s_{\mathrm{targ}}$ is catastrophically disrupted using the inverse of the mean free time:

\begin{equation}\label{single-rate}
\begin{split}
R_{\mathrm{coll,ind}}(s_{\mathrm{targ}},a) = & \, N(s_{\mathrm{targ}}) \, n(s_{\mathrm{imp}}) \\
& \times \sigma(s_{\mathrm{targ}},s_{\mathrm{imp}}) \, v_{\mathrm{rel}}(a)
\end{split}
\end{equation}

where $n( s_{imp})$ is the number density of the impactor planetesimal, $\sigma(s_{targ},s_{imp})$ is the gravitational focusing cross-section between the impactor and target. The gravitational focusing cross-section is defined as:

\begin{equation}
    \sigma(s_{targ},s_{imp}) = \sigma_{geo}(s_{targ},s_{imp}) (1+(\frac{v_{esc}}{v_{rel}(a)})^2)
\end{equation}

where $\sigma_{geo}(s_{targ},s_{imp})$ is the geometric cross section of the colliding bodies, $v_{esc}$ is the mutual escape velocity between the colliding bodies, and $v_{rel}(a)$ is the average relative velocity between the bodies.

Importantly, Equation \ref{single-rate} produces a collision rate of an individual target colliding with an individual impactor. For our purposes, we want to understand the rate of catastrophic collision for all bodies that are greater than or equal to a target of radius, $s_{targ}$ in the entire disk. To do this we modify Equation 2 to produce:


\begin{equation}\label{totalrate}
\begin{split}
R_{\mathrm{coll}}(s_{\mathrm{targ}},a) = & \sum_{s \ge s_{\mathrm{targ}}}^{\infty} \sum_{s_{\mathrm{imp,i}} \ge s_{\mathrm{imp}}}^{s} N(s)\,n(s_{\mathrm{imp,i}}) \\
& \times \sigma(s,s_{\mathrm{imp,i}})\,v_{\mathrm{rel}}(a).
\end{split}
\end{equation}

In this approach, we iterate through the targets with s $\geq s_{\mathrm{targ}}$ and the associating impactors, $ s_{\mathrm{imp,i}}$, that can catastrophically disrupt bodies of size $s$. From this, we can calculate a collision rate for each target/impactor combination by multiplying the individual collision rate by the total number of objects for each target/impactor combination. At the end, we have a disk wide catastrophic collision rate for objects $\geq s_{\mathrm{targ}}$.

To calculate the probability of a single or multiple collision events occurring in a certain period of time, we use Poisson statistics. We treat planetesimal collisions as Poisson processes, where collision events are independent of one another and, on scales of tens to hundreds of years, have a constant average collision rate. To calculate the probability that exactly $k$ collision events occur within a period of time, $t$, we use the probability mass function defined as:

\begin{equation}
    P_{pmf}(k) = \frac{e^{-(R_{coll}(s_{targ},a) \times t)} \times (R_{coll}(s_{targ},a) \times t)^{k}}{k!}
\end{equation}

where $R_{coll}(s_{targ},r)$ is in units of collisions per year and $t$ is in units of years. Furthermore, to calculate the probability of at least $k$ collision events occurring within a period of time $t$, we use the survival function defined as:

\begin{equation}
    P_{sf}(k) = 1 - \sum_{i=0}^{k}P_{pmf}(i)
    \label{tab:sf}
\end{equation}

Within the model we use Python library {\tt\string scipy.stats.poisson} \citep{Virtanen2020}. Specifically, we use {\tt\string scipy.stats.poisson.pmf} for the probability mass function and {\tt\string scipy.stats.poisson.sf} for the survival function.

\subsection{Fomalhaut Main Belt Configuration}\label{fomalhaut-config}

\begin{table*}
\centering
\begin{tabular}{lclc}
\hline\hline
\textbf{Parameter} & \textbf{Description} & \textbf{Value} & \textbf{Reference} \\
\hline
\multicolumn{4}{l}{\textbf{Stellar Parameters}} \\
\hline
$M_\star$ ($M_\odot$)  & Mass             & 1.92 $\pm$ 0.02 & \citet{Mamajek2012} \\
$t_{\star}$ (Myr)      & System Age       & 440 $\pm$ 40  & \citet{Mamajek2012} \\
$d_\star$ (pc)         & Distance        & 7.70 $\pm$ 0.03  & \citet{vanLeeuwen2007} \\
\hline
\multicolumn{4}{l}{\textbf{Belt Parameters}} \\
\hline
$a_{\rm belt}$ (au)      & Belt semi-major axis                       & 140 & Assumed \\
$\Delta a_{FWHM}$ (au)  & FWHM width of the belt        & 15.6     & \citet{Lovell2025} \\
$e_{f}$                      &  Forced belt eccentricity                          & 0.126 $\pm$ 0.001  & \citet{Lovell2025}  \\

$e_{p}$                      &  Free eccentricity of mm dust                          & 0.0389 $\pm$ 0.0015  & \citet{Lovell2025} \\
$h$                      & Vertical aspect ratio              & 0.016 $\pm$ 0.001    & \citet{Lovell2025} \\

$M_{dust,ALMA}$ $(M_{\oplus})$          & Dust mass observed with ALMA (s $\le$ 1 mm)             & 0.021 $\pm$ 0.001    & \citet{Lovell2025} \\

\hline
\end{tabular}
\caption{Adopted parameters for the Fomalhaut system. }
\label{tab:disk_properties}

\end{table*}

\begin{figure*}
    \centering
    \includegraphics[width=\linewidth]{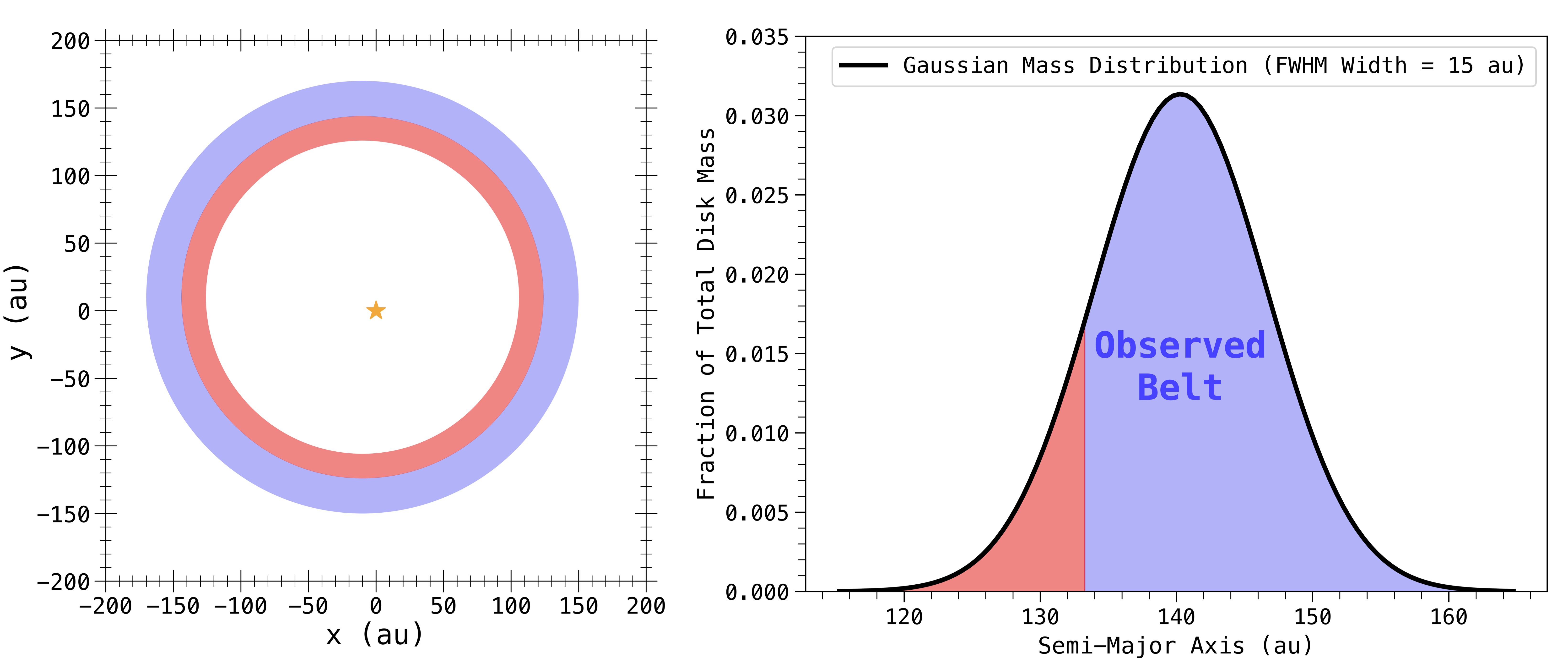}
    \caption{\textit{Left:} A schematic of the Fomalhaut main belt under our model. The blue region indicates the observed main belt, while the red region contains the low surface density, dynamically hot component of the main belt. \textit{Right:} The fraction of the belt mass in each region of the disk as a function of belt semi-major axis, which is governed by a Gaussian surface density distribution.}
    \label{fig:BeltSchematic}
\end{figure*}

We apply this model to the collisions observed in the Fomalhaut system \citep{Kalas2008,Kalas2025}. We consider the main planetesimal belt as the origin of the collisions seen in the Fomalhaut system due to their proximity to the main belt inner edge. We define the main belt as having an inner edge ($a_{min}$) of 115 au, an outer edge ($a_{max}$) of 165 au, the belt semi-major axis ($a_{belt}$) of 140 au, FWHM width ($\Delta a_{FWHM}$) of 15.6 au , and vertical aspect ratio ($h$) of 0.016. These values and corresponding references can be found in Table \ref{tab:disk_properties}. With the Gaussian surface density (see Section \ref{volume}), the low surface density region observed between the intermediate and main belt in Fomalhaut will contain < 5\% of the main belt mass.

For the collisions seen in Fomalhaut, we consider two components to the main belt. One component is the main belt where the dust has been observed from optical to mm wavelengths \citep[e.g.][]{Kalas2005,MacGregor2017,Gaspar2023} that has an inner edge at 134 au and an outer edge of 165 au. Furthermore, we assume the large planetesimals share the properties of the dust grains \citep[e.g.][]{Krivov2010,Pawellek2019}, which can be seen in Table \ref{tab:disk_properties}. Due to this assumption, the planetesimals in the observed main belt will have an average relative velocity of $\approx 4\%$ the Keplerian velocity in this model. We then consider a second component of the belt located interior to the observed belt edge located from 115 to 134 au, which is likely the origin location of the cs1 and cs2 collisions \citep[e.g.][]{Kalas2025}. We recognize that the cs2 collision event was found at a projected separation of 135 au \citep{Kalas2025}, but the high eccentricity and trajectory of the dust cloud, in addition to the nine year gap in which it could have occurred, indicates an origin likely interior to the observed inner edge, which we will assume moving forward. We further assume that planetesimals in the interior region have a higher relative velocity than the planetesimals in the observed belt. Stirring by an unseen interior planet in Fomalhaut can result in an increase in the perturbed planetesimal's eccentricity \citep[e.g.][]{Costa2024}. This is further supported by the derived relative velocity of $\approx$ 15\% the Keplerian velocity (upper limit for relative velocity of cs1 collision from \cite{Gaspar2020}) for the cs1 collision event, which we will use as the average relative velocity for this region in our model.

\subsection{Collisional Model Fitting}

Combining the collision model with the belt geometry and collisions observed in the Fomalhaut system, we outline the free parameters and physically-informed priors in the planetesimal model that are consistent with the two collisions interior of the observed belt.

We assume that the target planetesimal disrupted in the cs1 and cs2 collision events had a radius $\ge$ 100 km as found for cs1 by \cite{Gaspar2020}. Additionally, we run a model for the case that the target planetesimal have radii $\ge$ 30 km \citep{Kalas2025}. The analysis and results in Section \ref{results} will use the best-fit planetesimal population and collision rate values where the target radii are $\ge 100$ km. We will compare the results between the differing target planetesimal radii assumptions in Section \ref{30km}.

We fit the planetesimal population using {\tt\string emcee}, an Affine Invariant Markov Chain Monte Carlo (MCMC) Ensemble Sampler \citep{emcee1,emcee2}. The model will fit six parameters: the joint collisional cascade power-law index spanning both the strength and gravity dominated regime ($\alpha_{12}$), the power-law index governing the primordial planetesimal population ($\alpha_3$), the total mass of the belt ($M_{belt}$), the largest planetesimal present in the belt population ($s_{max}$), the break point where the planetesimal population transitions from the collisional evolved bodies to the primordial population ($s_{break}$), and the primordial population depletion parameter ($J$).

In the fit we chose to combine $\alpha_1$ and $\alpha_2$ into a single parameter, $\alpha_{12}$ to avoid individual, degenerate solutions to $\alpha_1$ and $\alpha_2$. We link the power-law indices setting $\alpha_2 = \alpha_1 + 0.5$. This preserves the difference between the two power law indices in their ideal cases (see Section \ref{dist}). We recognize that the two parameters do not require a difference of 0.5, and note this as a simplifying assumption of the model.

For a given set of the six parameters just described, our model calculates the total collision rate for a given region of the belt. Since the observed collisions likely originated interior to the observed belt inner edge, the collision rate is output for belt semi-major axes < 134 au. The collisions rate, being a random Poisson process, is then used, along with the period of observations ($t \sim 20$ yrs) and the number of collisions observed ($k=2$) to determine the the likelihood of the set of parameters to explain the observed collisions. Finally, the samples for the posterior distribution are determined using the logarithm of the probability mass function described in Section~\ref{model} and the priors.

In the planetesimal model fit since the collisions occur between large planetesimal bodies, the inferred dust mass from ALMA observations will be used to inform the power-law indices governing small grains and sub-km planetesimals through the $\alpha_{12}$ parameter. The final model will have a dust mass that is consistent with observations.

The age of the Fomalhaut system will also place an important constraint on the planetesimal population fit. Specifically in a primordial planetesimal population that is being collisionally disrupted over time, the transition point between the collisionally evolved and primordial population should occur roughly where the individual collision timescale equals the age of the system ($1/R_{coll,ind}(s) \sim t_{age}$; \cite{Wyatt2011,Lohne2008,Pearce2025}). In the case of Fomalhaut, that translates to the individual collision timescale of planetesimals of size $s_{break}$ being 440 $\pm$ 40 Myr \citep{Mamajek2012}. Since the planetesimals are assumed to originate from a common population but with different relative velocities, we determine a common $s_{break}$ by taking the average individual collision timescale throughout the entire disk.

\begin{table*}

\centering

\begin{tabular}{l|c|c|c}
\hline\hline
\textbf{Best-Fit Parameters} & \textbf{Description} & \textbf{Values} ($s_{targ} \ge 100$ km) & \textbf{Values} ($s_{targ} \ge 30$ km)  \\
\hline
$\alpha_{12}$          & Joint first and second power law index & $-3.65^{+0.01}_{-0.03}$ & $-3.78^{+0.03}_{-0.03}$ \\
$\alpha_3$             & Third power law index  &$-5.05^{+1.63}_{-1.85}$ & $-4.95^{+1.77}_{-1.85}$ \\
$M_{belt}$ [$M_{\oplus}$] & Total belt mass & $248.68^{+110.39}_{-51.08}$ &  $11.82^{+8.21}_{-4.67}$\\
$s_{break}$ [km]       & Evolved-primordial planetesimal transition radius & $114.82^{+29.72}_{-10.10}$ & $33.11^{+18.17}_{-6.20}$ \\
$s_{max}$ [km]         & Maximum planetesimal radius in population &$380.19^{+643.10}_{-202.36}$ & $131.82^{+248.36}_{-78.12}$  \\
log($J$)                 & Primordial depletion parameter & $-0.62^{+0.74}_{-0.86}$ & $0.14^{+0.61}_{-0.88}$ \\

\hline

\end{tabular}
\caption{Best-fit model results for two collisions with target planetesimal for the case that the target planetesimals were of radii $\ge$ 100 km and $\ge 30$ km in Fomalhaut's main belt. The lower and upper bounds are given by the 16th and 84th percentile of our posterior distribution, respectively. }
\label{results_vertical}
\end{table*}

\section{Results}\label{results}

In this section, we use the best-fit planetesimal population and belt model to derive collision rates for planetesimals in the Fomalhaut debris belt. From these collisions, we derive the probability of observing collisions for a range of planetesimals, including those similar to the observed collisions that have been observed thus far.

\begin{table*}[t]
\centering
\label{tab:results_derived}
\begin{tabular}{l|c|c|c}
\hline\hline
\textbf{Derived Parameters} & \textbf{Description} & \textbf{Values} ($s_{targ} \ge 100$ km) & \textbf{Values} ($s_{targ} \ge 30$ km) \\
\hline
$R_{coll}(s \ge s_{targ})$  [yr$^{-1}$]  & Catastrophic coll. rate for $s \ge s_{targ}$ within $ 134$ au  & $0.086^{+0.067}_{-0.048}$ & $0.081^{+0.075}_{-0.037}$ \\
$t_{coll,ind}(s_{break})$ [Myr] & Individual coll. timescale for $s = s_{break}$  &$450^{+47}_{-45}$ &  $445^{+50}_{-48}$\\
$M_{dust}$ [$M_{\oplus}$] & Dust mass for $s < 1$ cm & $0.021^{+0.001}_{-0.001}$ & $0.021^{+0.001}_{-0.001}$ \\

\hline
\end{tabular}
\caption{Derived parameters from best-fit model results}
\label{derived-vals}

\end{table*}

\subsection{Best-Fit Planetesimal Population}

\begin{figure}
    \centering
    \includegraphics[width=\linewidth]{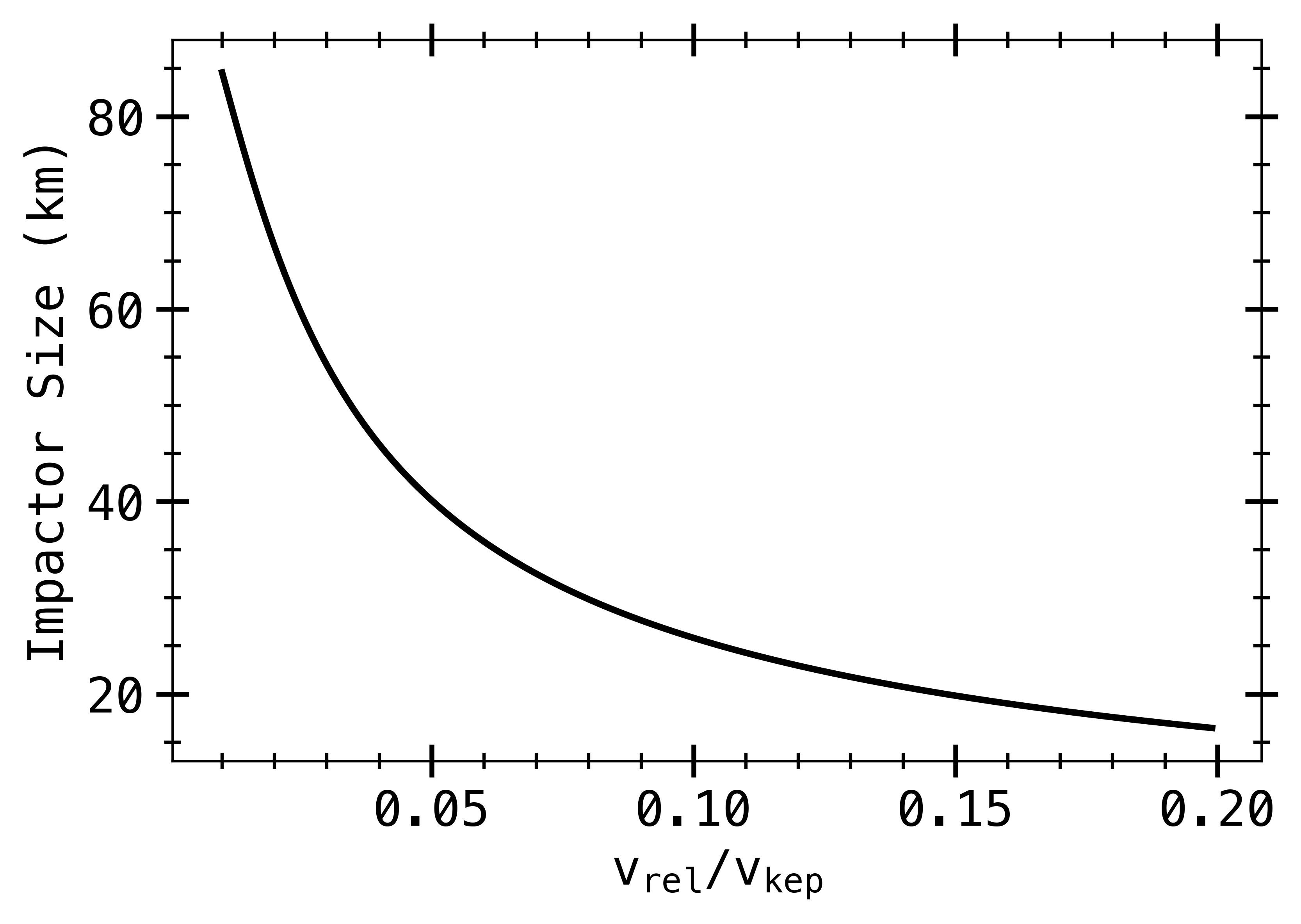}
    \caption{Minimum impactor radius needed to catastrophically disrupt a planetesimal with a radius of 100 km as a function of the relative velocity of the impact.}
    \label{impactor}
\end{figure}

\begin{figure*}
    \centering
    \includegraphics[width=\linewidth]{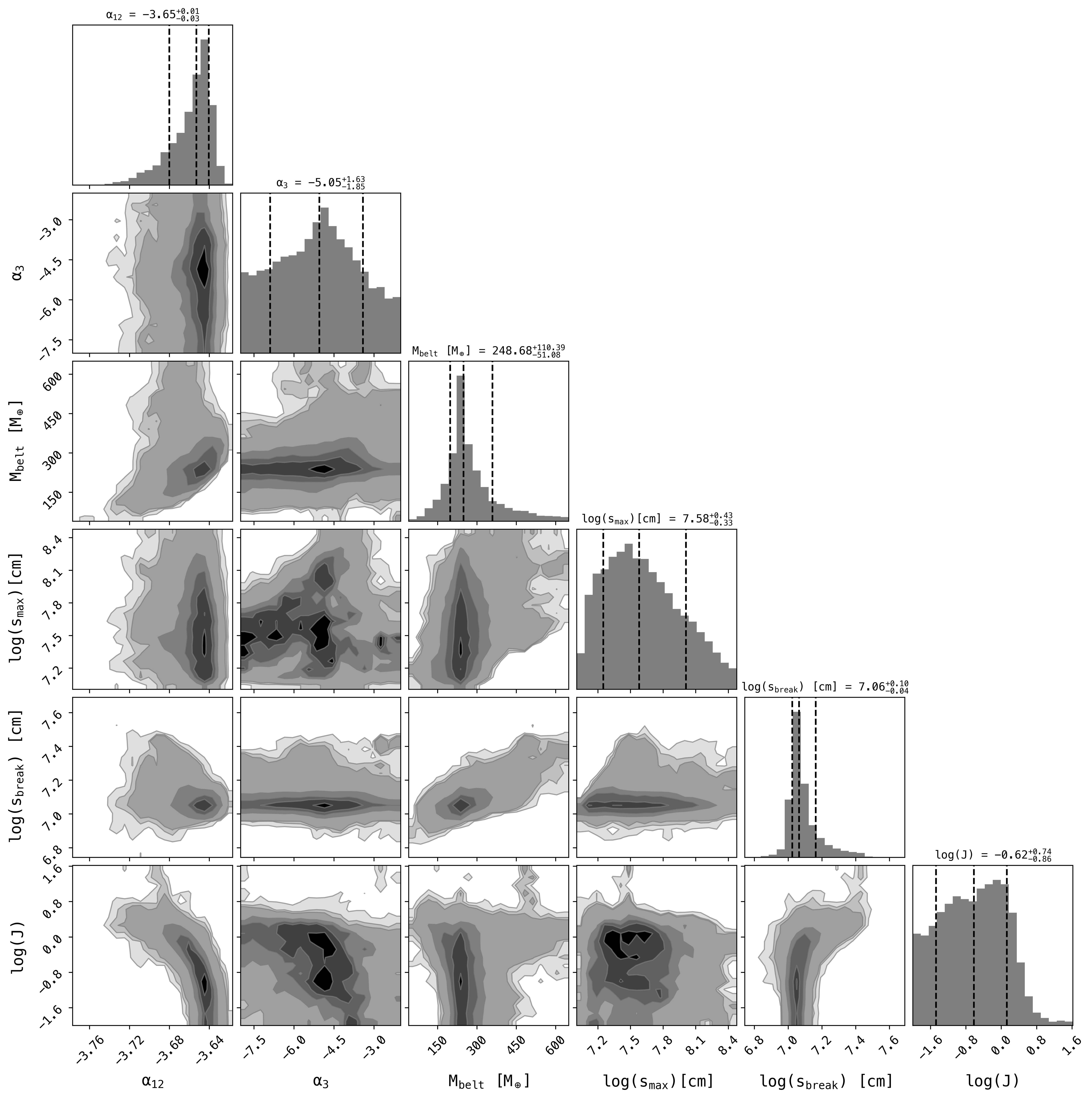}
    \caption{The posterior distribution of the {\tt\string emcee} chain of the six parameters fit in our model for collisional progenitors with radii $\ge$ 100 km. The model is fit with 50 walkers, 10000 steps, and 1000 burn-in steps. The dashed lines indicate the 16th, 50th, and 84th percentiles, respectively.  }
    \label{posterior100}
\end{figure*}

\begin{figure*}
    \centering
    \includegraphics[width=\linewidth]{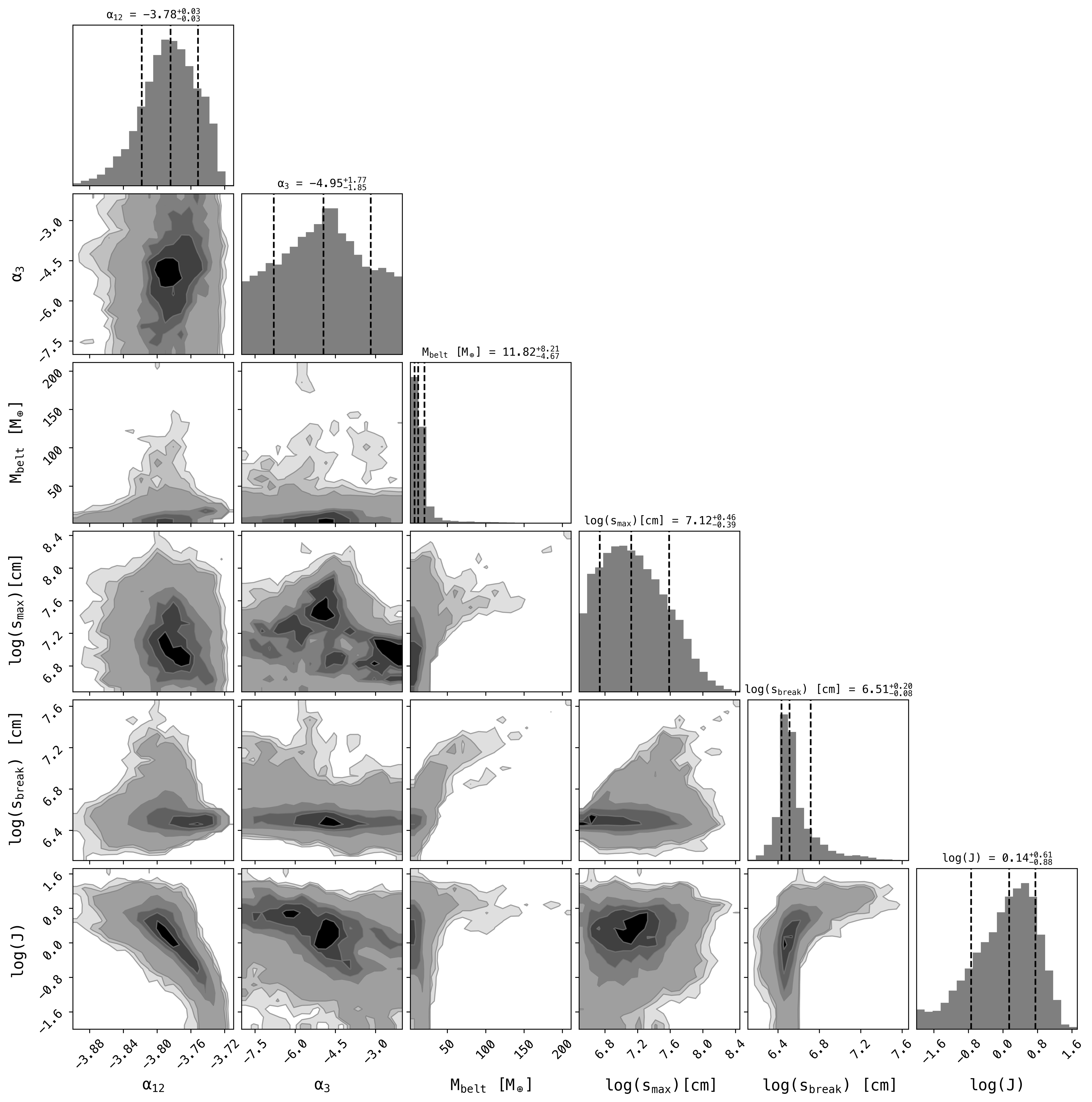}
    \caption{The posterior distribution of the {\tt\string emcee} chain of the six parameters fit in our model for collisional progenitors with radii $\ge$ 30 km. The model is fit with 50 walkers, 10000 steps, and 1000 burn-in steps. The dashed lines indicate the 16th, 50th, and 84th percentiles, respectively.  }
    \label{posterior30}
\end{figure*}

We fit the underlying planetesimal population to find population parameters consistent with the two observed collision events, in addition to the dust mass and system age priors, in the last 20 years where both of the target planetesimals are $\ge$ 100 km in radius. Due to the assumed higher relative velocities in the region interior to the belt observed inner edge, the minimum impactor radius that can disrupt a 100 km radius planetesimal is $\sim$20 km, compared to $\sim$ 60 km in the observed main belt (see Figure \ref{impactor}).The best-fit model posterior distributions can be found in Figures \ref{posterior100} and \ref{posterior30}, and the best fit parameters can be found in  Table \ref{results_vertical}.

\begin{figure*}
    \centering
    \includegraphics[width=\linewidth]{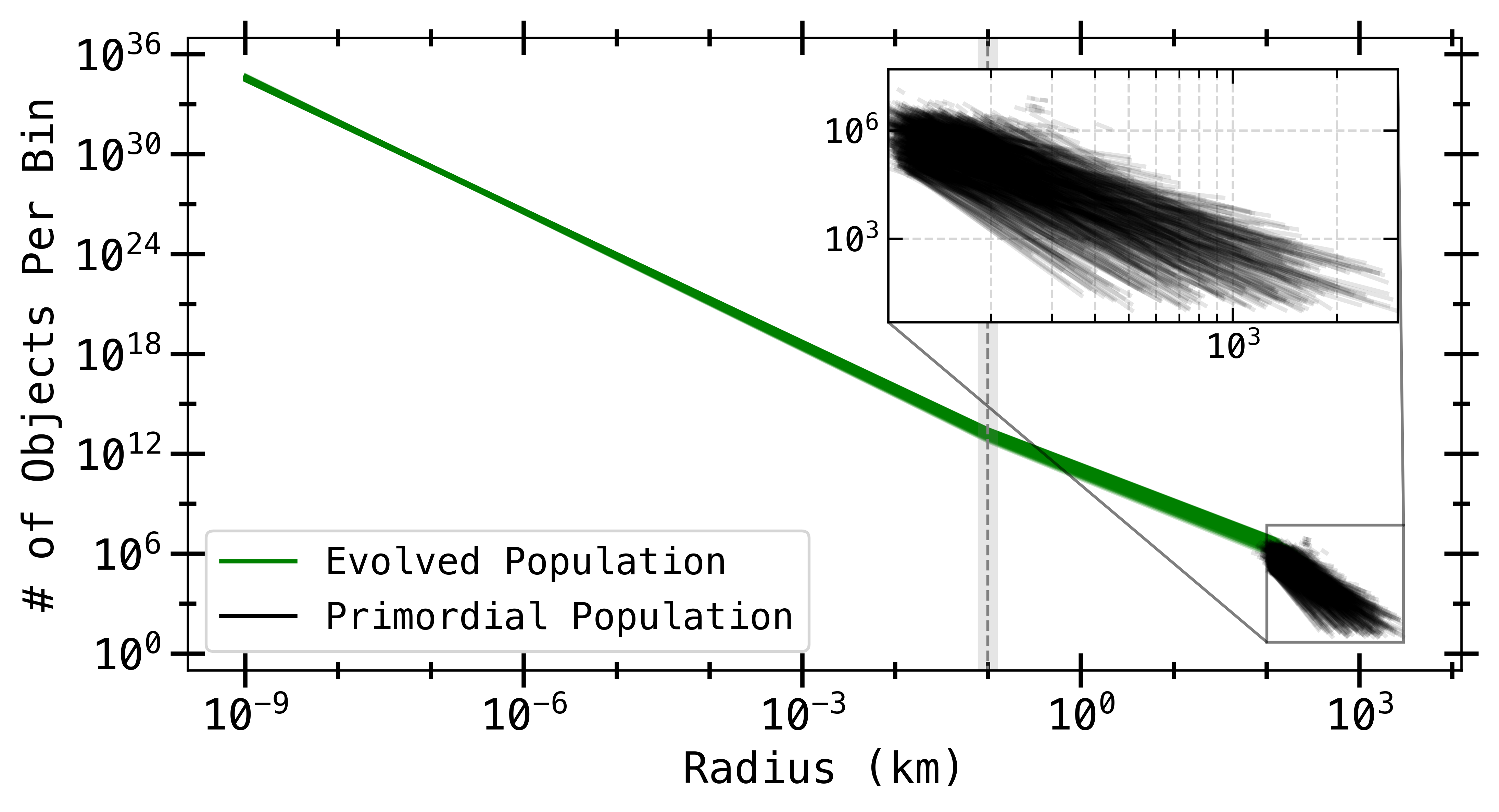}
    \caption{1000 posterior draws from the best-fit planetesimal population. The green lines indicate the evolved planetesimal population, where the planetesimals have been collisionally processed and follow a power-law consistent with a collisional cascade. The gray dashed line indicates the point where the the collisional cascade transitions from particles whose cohesion is mainly from material strength to cohesion from self-gravity \citep[e.g.][]{Pan2005}. The model jointly fits the strength and gravity power law indices in $\alpha_{12}$. The black lines indicate the best-fit primordial planetesimal population, which is made up of planetesimals that have individual collision timescales that are longer than the age of the Fomalhaut system. The planetesimal radius where the evolved population transitions to the primordial population is $s_{break}$.}
    \label{fig:pop}
\end{figure*}

From the best-fit model, we find a total belt mass between 200-360 $M_{\oplus}$ with the most likely mass of $\sim 250 \text{ }M_{\oplus}$. The planetesimal population transitions from the collisionally evolved to the primordial population occurs at a radius of $115^{+30}_{-10}$ km. Additionally, we are able place loose constraints on the primordial planetesimal population (i.e. $s>s_{break}$). We find the largest planetesimal present in the belt is between 180-1020 km in radius with most likely maximum planetesimal radius of 380 km. We also place mild constraints on the planetesimal depletion factor, $J$, with the primordial population being either continuous with the collisionally evolved population or slightly depleted. Finally, the power law index governing the primordial population $\alpha_3$ peaks at -5, indicating a steep primordial population, but shallower slopes are not entirely ruled out (i.e., $\le -3$). The best-fit planetesimal population from our model fit can be found in Figure \ref{fig:pop}.

\subsection{Collision Rates Interior to the Main Belt Inner Edge}

\begin{figure*}
    \centering
    \includegraphics[width=\linewidth]{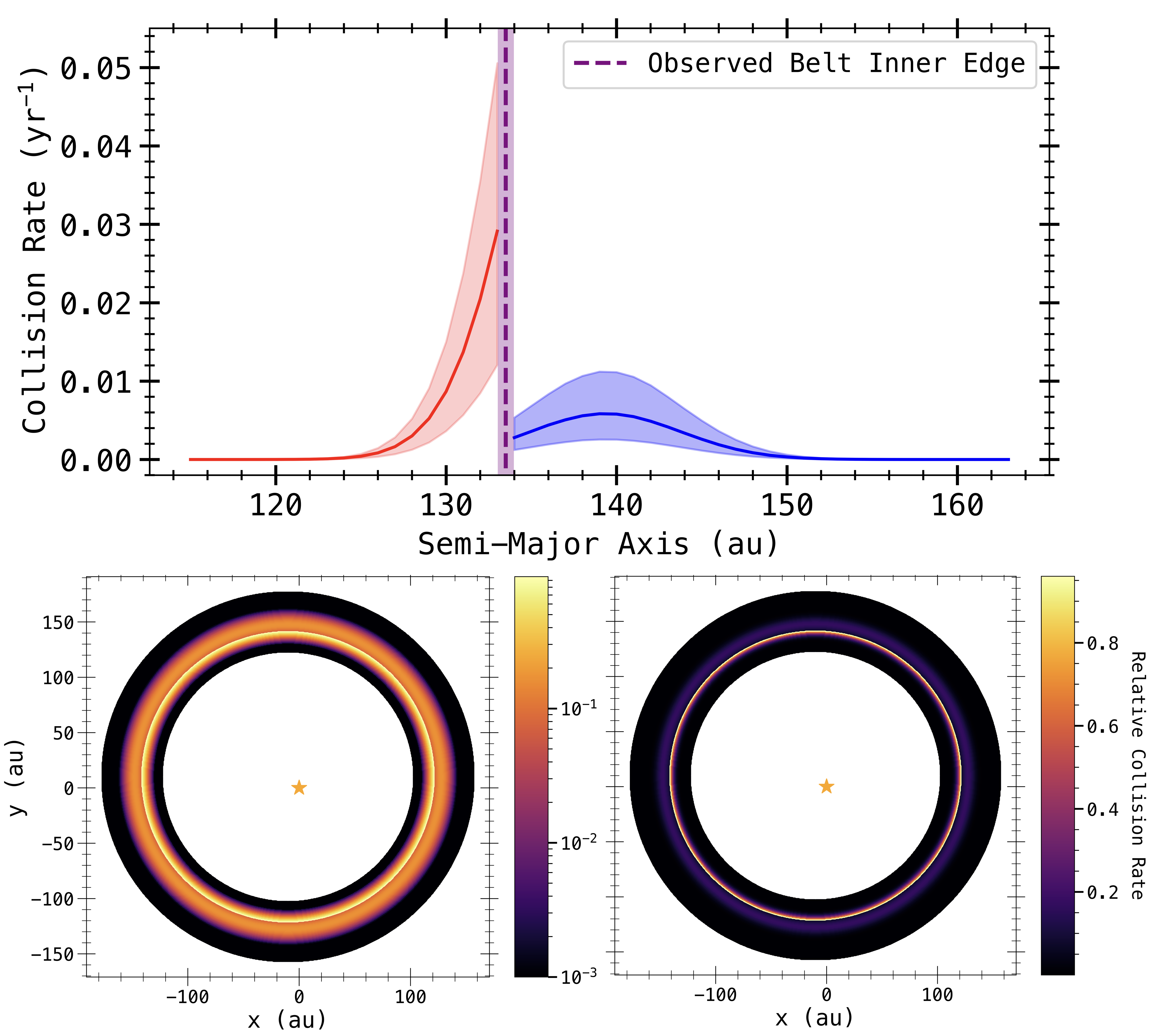}
    \caption{\textit{Top:} The catastrophic collision rate for target planetesimals $\ge$ 100 km in radius as a function of belt semi-major axis. The blue region indicates the observed Fomalhaut main belt, and the red region indicates the region where Fomalhaut cs1 and cs2 originated, which is within observed belt inner edge. We find the highest collision rate at the observed inner edge of the Fomalhaut main belt. The shaded regions indicate the 16th and 84th percentiles of the posterior distribution. We find a very low collision rate interior to 125 au due to the low surface density of planetesimals in that region. We find a non-negligible collision rate in the observed main belt region. \textit{Bottom:} Belt schematics outlining the relative collision rate for target planetesimals $\ge$ 100 km in radius. The two images plot the same data, with the left on a logarithmic scale and the right on a linear scale. The brightest band in both images is the location of observed belt inner edge and the location where the collision rate peaks. This is coincident with the discovery location of the most recent collision, Fomalhaut cs2. }
    \label{fig:coll_rate}
\end{figure*}

In the region of the disk interior of the inner edge of the Fomalhaut belt, we derive a total catastrophic collision rate for target planetesimals with radii $\ge 100$ km of $0.086^{+0.067}_{-0.048}$ collision events per year (see Table \ref{derived-vals}). The collision rate, however, is not uniform throughout this region, with the collision rate peaking at the inner edge of the observed belt, with a sharp drop away from the interior to the observed inner edge (see Figure \ref{fig:coll_rate}). The location where the collision rate peaks encompasses the discovery location of Fomalhaut cs2, while the original Fomalhaut cs1 was discovered 10 au interior of the observed belt inner edge at a projected separation of 120 au \citep{Kalas2008}.

\begin{figure}
    \centering
    \includegraphics[width=\linewidth]{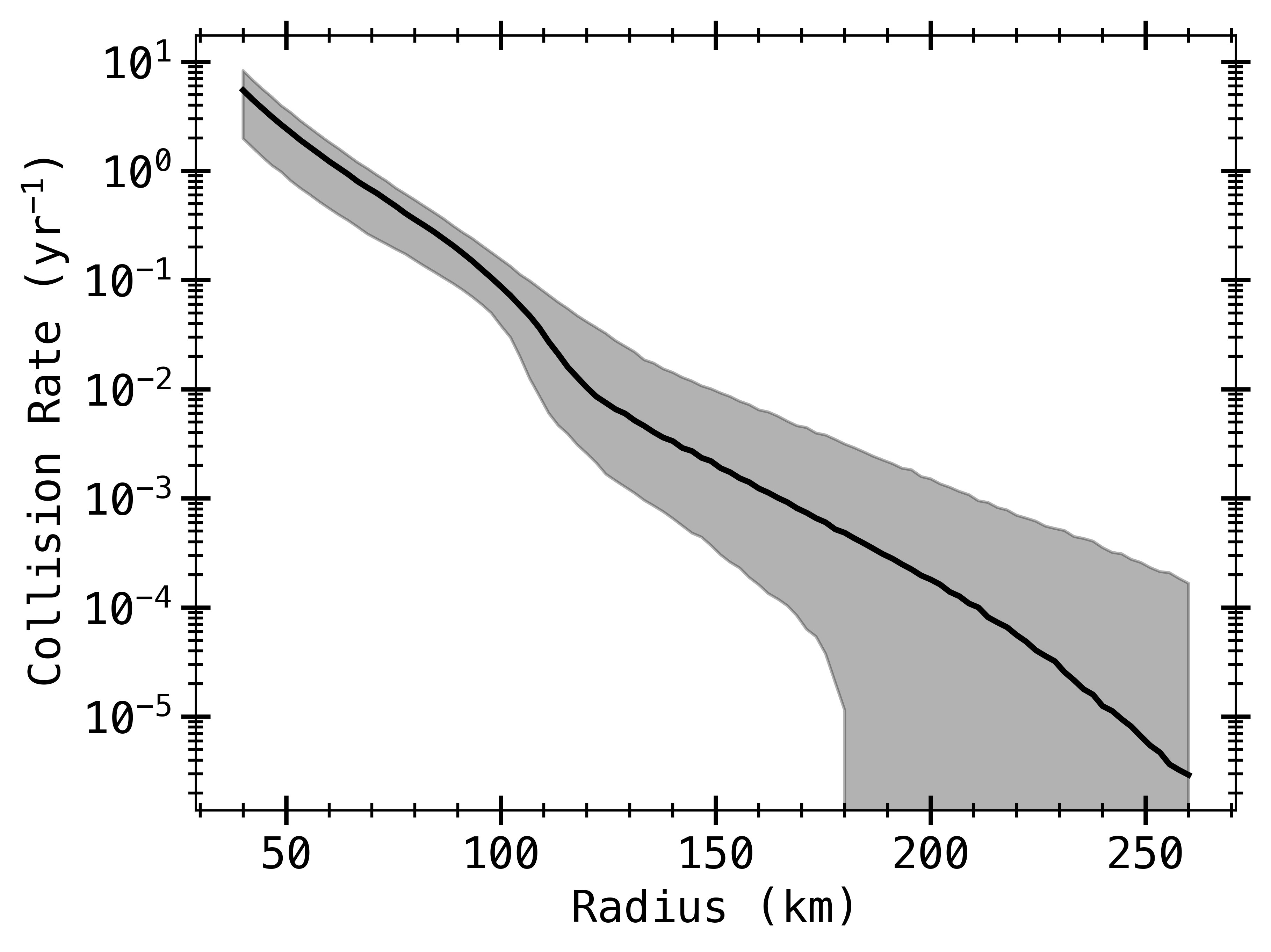}
    \caption{The total collision rate in the dynamically hot region interior of the belt observed inner edge as a function of target planetesimal radius, assuming the cs1 and cs2 events were caused by target progenitors $\ge$ 100 km in radius. For a given radius on the x-axis, the collision rate is the collision rate for target planetesimals greater than or equal to that radius. For example, the corresponding collision rate for a planetesimal with a radius of 50 km on the x-axis is the collision rate for target planetesimals with radii $\ge$ 50 km.}
    \label{fig:all-radii}
\end{figure}

We then apply the best-fit planetesimal population and collision model to target planetesimals with radii $\ge 50$ km and $\ge 200$ km, finding collision rates of $2.29^{+1.36}_{-1.41}$ and $0.0021^{+0.0010}_{-0.0020}$ collision events per year, respectively (see Figure \ref{fig:all-radii}). This implies that, under this model, within the same time period Fomalhaut cs1 and cs2 were observed, there were between 18-73 collision events between smaller planetesimals. Conversely under this model catastrophic planetesimal collision events with target radii $\ge 200$ km are quite rare but still plausible to be observed, occurring once every 300-1000 years under these best-fit belt conditions.

\subsection{Collision Rates Within the Observed Main Belt}

 Within the observed main belt, we find a total catastrophic collision rate for target planetesimals with radii $\ge 100$ km of $0.057^{+0.043}_{-0.032}$ collision events per year. This is calculated by summing the collision rate in the observed belt region, as seen in the top panel of Figure \ref{fig:coll_rate}. This implies that under this model at least one collision event comparable to the Fomalhaut cs1 and cs2 collision events likely occurred in the observed main belt in the last 20 years. Although the Fomalhaut cs1 and cs2 collision events likely originated interior to the belt inner edge, we find a comparable collision rate within the observed belt as well. This is due to the high surface density in the observed main belt even though the average relative velocities between these regions differ by a factor of $\approx$ 4 (see Section \ref{fomalhaut-config}).

\subsection{Next Collision in Fomalhaut}

\begin{figure*}
    \centering
    \includegraphics[width=\linewidth]{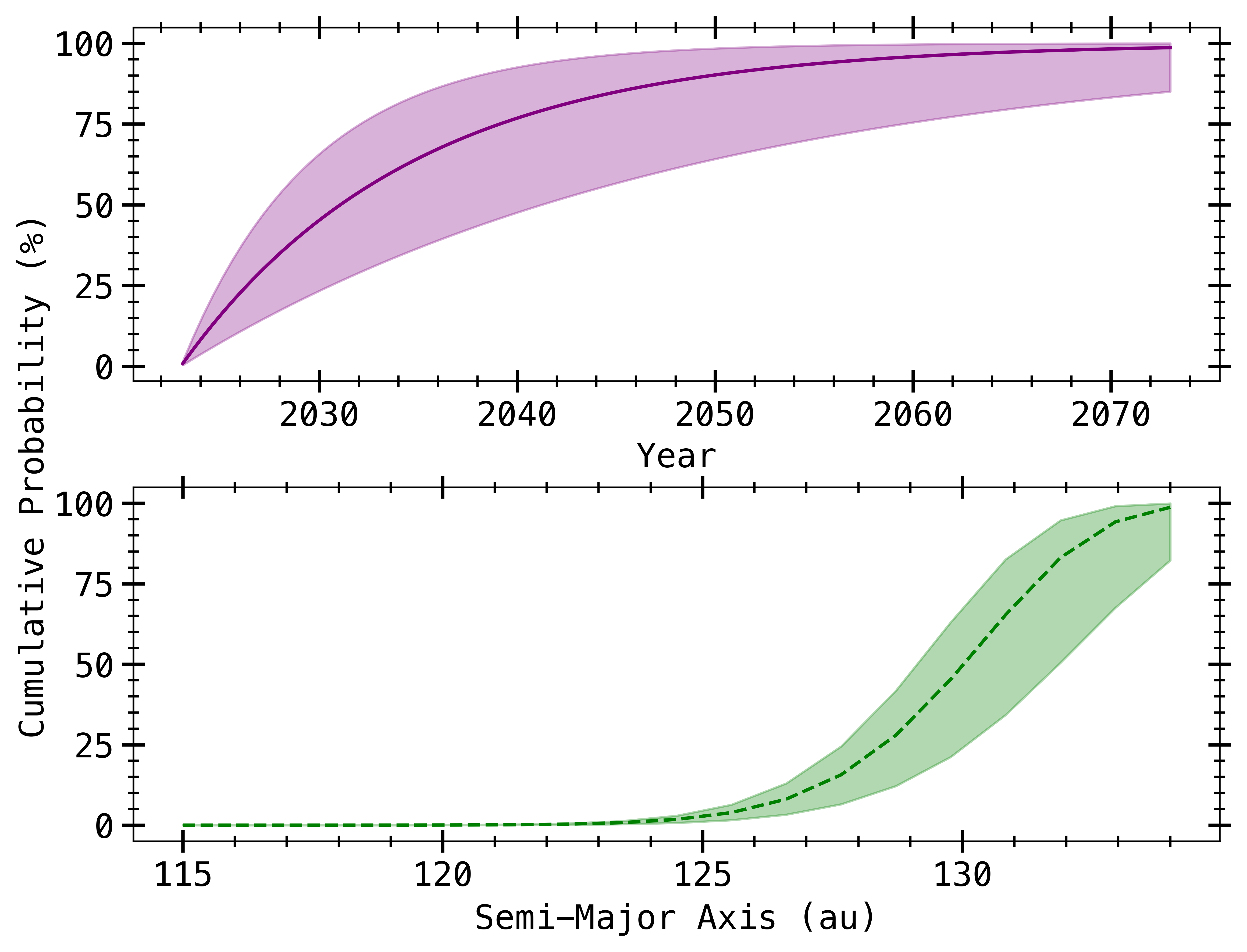}
    \caption{\textit{Top:} The cumulative probability of the next collision event (a hypothetical Fomalhaut cs3) where the target planetesimal radius is $\ge$ 100 km in radius as a function of collision year under the conditions of our model (see Table \ref{tab:results_derived}). The shaded regions indicate the 16th and 84th percentiles of the posterior distribution. We find a 50\% chance of the next collision occurring in 2027 at earliest and 2041 at the latest. \textit{Bottom:}  The cumulative probability of the next collision event (Fomalhaut cs3) where the target planetesimal radius is $\ge$ 100 km in radius as a function of belt semi-major axis.}
    \label{fig:cumulative}
\end{figure*}

Using the best-fit collision rate for the planetesimal population in the Fomalhaut belts, we predict when and where we expect a Fomalhaut cs3 to occur in the future (see Figure \ref{fig:cumulative}) using Equation \ref{tab:sf}. Under this model, we predict there is a 50\% chance that the next comparable collision to cs1 and cs2 will occur by calendar year $2031^{+10}_{-4}$ CE, 75\% by $2039^{+20}_{-7}$ CE, and 90\% by $2049^{+23}_{-11}$ CE.

For the collision location under this model, the probability of the next collision occurring within 125 au is $1.7^{+0.9}_{-1.2}\%$, within 130 au is $42.7^{+19.3}_{-22.1}\%$, and within 135 au is $97.5^{+2.5}_{-19.8}\%$. We note that probabilities for the collision location are independent of the predicted year for the next collision. From this, we would expect the next collision to occur in a similar location to that of the Fomalhaut cs2 event, while the probability of it being found near the original Fomalhaut cs1 event is unlikely under this model.

\section{Discussion}

\subsection{Location of Future Collisions and the Influence of Model Assumptions}

The location of the Fomalhaut cs1 collision event remains uncertain, with the discovery image placing it at 120 au if it is coplanar with the disk \citep{Kalas2008,Kalas2013}, while other analyses only place constraints that the collision location is < 150 au \citep{Gaspar2020}. In our model, we use a simplistic, piecewise approach, where the average collision velocity in the observed main belt is 4\% the Keplerian velocity and the average collision velocity interior to the observed inner edge is 15\% the Keplerian velocity. Furthermore, we assume a Gaussian surface density distribution centered at 140 au and falls off as a function of distance from the peak. Due to the surface density assumption, we expect belt semi-major axes < 125 au to be an unlikely location for the next collision be found. In contrast, this model predicts the highest collision rate region to be at the observed belt inner edge due to the combination of high relative velocities and higher surface density (see Figure \ref{fig:coll_rate}). However, the discovery of a future collision at these separations (< 125 au) is likely higher than the model prediction. Since the only constraint on the relative velocity of the planetesimals interior to the belt inner edge comes from Fomalhaut cs1 \citep{Gaspar2020}, the true surface density of planetesimals in this region may not continue to follow the Gaussian distribution as we have assumed (see Figure \ref{fig:BeltSchematic}) and may become broadened due to a higher relative velocity dispersion. However, we do not expect a variation in the shape of the surface density to affect the total collision rate in the region interior to the belt edge and the best-fit planetesimal properties, as long as the assumed average relative velocity in the hot region from the cs1 event is not anomalously small or large compared to the true distribution and the total collision rate in the hot region is constant, even if the shape of surface density distribution varies. Small variations should mainly shift the collision probability density from being concentrated at inner edge of the observed belt (as seen in the bottom panel of Figure \ref{fig:cumulative}) to regions more interior. To test this, we have run an additional 100 km progenitor simulation with a smoothed/plateau surface density in the region where the collisions have been observed, and we find all best-fit parameters are consistent with our Gaussian surface density version within uncertainty (see Appendix \ref{appendix-A}). Additionally, this also holds true for when we expect the next collision to occur in Fomalhaut. The only result to significantly change is where the collision is expected to occur within the belt inner edge, with there now being a $\ge$ 50$\%$ chance the collision occurs within 125 au as opposed to $\le$ 1$\%$ in the Gaussian tail version. As the expansion and evolution of Fomalhaut cs2 is monitored over the coming years and as additional collisions are discovered, it will provide further constraints on the relative velocities and surface density distribution between colliding planetesimals in this region.

\subsection{Implications of a Non-Detection of an Additional Collision in the Observed Main Belt}

A main result from our best-fit planetesimal population is that at least one collision event, comparable to the cs1 and cs2 events, likely occurred in the low free eccentricity, observed main belt of Fomalhaut ($0.057^{+0.043}_{-0.032}$ collision events per year or 0.5-2 collision events in 20 years). Here we will explore how such a collision could be detected and how the non-detection of this collision will further inform the planetesimal and belt model.

Assuming that the observed Fomalhaut main belt is optically thin, a multi-epoch analysis of archival HST images of Fomalhaut, similar to the analysis done for $\beta$ Pictoris in \cite{Apai2015} and \cite{Avsar2024}, could uncover localized, high signal-to-noise changes in surface brightness from the excess dust created from the hypothetical collision event. This approach would require multi-epoch analyses to be sensitive to percent-level surface brightness changes caused by a 100 km radius target planetesimal being catastrophically disrupted. One scenario where a collision has occurred in the observed main belt, but avoid detection is if it occurs within the regions of the Fomalhaut belt that is heavily obscured by PSF-subtraction residuals or lie outside of the field-of-view of HST \citep{Kalas2025}. While this scenario is possible, we believe this analysis will be a crucial test of this model, and we plan to pursue this analysis in a separate study in the near future.

In the case that such a collision did not take place, this will place important constraints on the total mass of the belt and the dynamical state of the planetesimals that are interior to the observed edge of the main belt that are likely responsible for the cs1 and cs2 events. A major limiting assumption in our model, is that the planetesimals in this dynamically hot region interior to the main belt collide at an average relative velocity of 15\% the Keplerian velocity. Furthermore, the relative velocity plays a large role in the collision rate of planetesimals under this model (see Equation \ref{totalrate}). The collision rate in turn dictates the mass needed in the belt to reproduce the observed collision events (cs1 and cs2) within the period of time in which they were observed for a given relative velocity. Additionally, the relative velocity in the main belt is determined by the belt width and is considered constant. Since we assume the relative velocity in the main belt is a relatively well constrained observable via the belt width, the non-detection of the additional collision inferred from the best-fit planetesimal population in the observed belt would imply that the model is overestimating the disk mass and underestimating the average relative velocity in the region where cs1 and cs2 occurred. In other words, searching for an additional collision event in the observed main belt will provide critical constraints on the dynamical state interior to the inner edge of the main belt, which will in turn tighten the constraints on the allowed total belt masses under this model.

\subsection{Imaging Fomalhaut with Next Generation Observatories}

Under our best-fit planetesimal model for the Fomalhaut system, we estimate between $\sim$ 20-70 collisions with target planetesimals with radii $\ge$ 50 km to have occurred in the last 20 years in the region of the disk where cs1 and cs2 likely originated. We will explore the implications of the non-detections with HST and JWST and how next generation observatories can provide further constraints for our model. While a full radiative transfer analysis of the dust production and evolution is beyond the scope of this work, simple geometric scaling suggests that these events should be readily detectable by future missions. Assuming the brightness of the resulting dust cloud scales with the cross-sectional area of the progenitor \citep[e.g.][]{Kenyon&Bromley2004}, a catastrophic disruption of a 50 km radius body would produce a dust cloud initially $\sim$25\% as bright as Fomalhaut cs1 (assuming a $\sim$100 km progenitor for cs1). Such an object would fall below the detection threshold of current HST coronagraphic imaging, explaining the lack of additional detections to date. However, the several order-of-magnitude contrast increase with next generation space-based observatories, like HWO, should be sensitive to the remnants of such collisions according to our brightness scaling \citep[e.g.][]{Debes2019,Decadal2020}.

If the expected $\sim$ 20-70 smaller collisions events are occurring in the Fomalhaut disk, we would expect the appearance of multiple disrupted planetesimals assuming regular monitoring of the system is occurring, as seen with Fomalhaut cs1 and cs2. More interestingly, if many smaller collisions are continuously occurring in the region interior to the main belt inner edge, then we expect the dust produced from these collisions will shear into a clumpy ring-like feature in between the main belt and intermediate belt \citep[e.g.][]{Kenyon&Bromley2004}.

The non-detection of such dimmer point sources or ring-like features between the intermediate and main belt can have two different implications. The non-detection of smaller collisions, with continual major collisions (i.e. if we find Fomalhaut cs3) could imply a steeper planetesimal distribution in this region compared to the main belt, resulting in fewer smaller planetesimals compared to larger ones. However, if a cs3-like collision does not occur in the next $\approx$20-30 years and the dimmer point sources or ring-like features are not detected with future observatories, this would imply this region of the Fomalhaut belt is dynamically colder and/or less massive than assumed in our model. Furthermore, this would imply that the detection of the cs1 and cs2 collisions could be stochastic outliers when compared to the true time-averaged collision rate.

\subsection{Comparing Collision Progenitor Size Estimates}\label{30km}

There are currently two estimates for the radii of colliding progenitors that are responsible for the Fomalhaut cs1 and cs2 events. Dynamical evolution modeling by \cite{Gaspar2020} suggests that the colliding progenitor must have a radius of at least 100 km, while \cite{Kalas2025} argue that the progenitors are likely $\sim$30 km in radius. While 100 km progenitors can explain the amount of dust created in the cs1 event, \cite{Kalas2025} argue that, if such large planetesimal collisions are typical, a the amount of dust created over the age of the Fomalhaut system would exceed the steady state mass loss rate. Therefore, they suggest that a 30 km radius planetesimal could be responsible for the cs1 and cs2 events. Such lower-mass planetesimals would be consistent with the steady-state dust production late as well as the dust produced by the observed cs1 event.

For this reason, we run our model for both progenitor sizes, and the results are available in Tables \ref{results_vertical} and \ref{tab:results_derived}. The key difference between the two results is the resulting total belt mass. A target progenitor of radius $\ge$ 100 km yields a belt mass ranging between 200-360 $M_{\oplus}$, while the $\ge$ 30 km case yields a belt mass roughly between 5-20 $M_{\oplus}$. One simple check is to see whether the 200-360 $M_{\oplus}$ estimate exceeds the total solid mass reservoir in Fomalhaut's protoplanetary disk. Assuming a dust-to-gas ratio of 0.01 and a disk-to-stellar mass ratio of 0.1 in Fomalhaut's protoplanetary disk\citep[e.g.][]{han2023}, the initial mass of the Fomalhaut planetesimal belt should be no more than 650 $M_{\oplus}$, which is consistent with either case of progenitor size. It is difficult to further distinguish between these two possible progenitor size regimes due to the wide range of the total belt masses derived from previous works \citep{Kalas2005,Acke2012,Gaspar2020,Krivov&Wyatt2021,Pearce2025}. For example, \citet{Krivov&Wyatt2021} provide a belt mass estimate of between 1.8-360 $M_{\oplus}$, encompassing the mass the range derived for both 100 km and 30 km progenitor sizes.

We believe one way to distinguish between the progenitor sizes responsible for cs1 and cs2 will be by breaking the degeneracies in the Fomalhaut belt mass, potentially through dynamical studies like that of \cite{Sefilian2025}. From the results of our model, a large belt mass (> 100 $M_{\oplus}$) would favor a progenitor size closer to 100 km as found by \cite{Gaspar2020} and a small belt mass (< 100 $M_{\oplus}$) would favor smaller progenitor sizes as found by \cite{Kalas2025}.

Importantly, we note that the collision rate estimates and predictions provided in Figures \ref{fig:coll_rate} and \ref{fig:cumulative} apply for both progenitor cases as both models are able to match the observed detection rates as seen in the first row of Table \ref{tab:results_derived}.

\subsection{Planetesimals Interior to the Observed Belt Inner Edge}

In our model, we make the assumption that the relative velocity of 15\% the Keplerian velocity for the dynamically hot region of the disk interior to the observed inner edge of the main belt of Fomalhaut. Here we explore a back of the envelope calculation to determine the maximum eccentricity the planetesimals in this population can occupy under certain simplifying assumptions. First, we assume that the planetesimals in this region originate from the main planetesimal belt starting at 134 au. Secondly, we assume that the orbits of these planetesimals do not cross the intermediate belt and are confined in orbits between the intermediate and main belt. Finally, we assume that the apocenter of the new orbit of a perturbed planetesimal is the inner edge of the observed main belt at 134 au.

With an orbital apocenter of 134 au and the smallest possible pericenter of of 104 au yields a maximum orbital eccentricity of approximately 0.13. This yields a relative velocity of approximately 14\% of the Keplerian velocity between planetesimals in this population, assuming negligible inclination effects, and is roughly consistent with our choice of 15\% the Keplerian velocity. Furthermore, this origin and orbit would result in the most collisions at apocenter, as the planetesimals in this orbit will spend more of their orbital period near apocenter compared to pericenter. This is consistent with our model prediction and the location of the new cs2 event \citep{Kalas2025}.

A plausible pathway for populating the interior region is through gravitational scattering or secular excitation of eccentricities by a planet near the belt's inner edge, which can drive planetesimals onto orbits with pericenters interior to the main belt \citep[e.g.][]{Wyatt2008}. However, we emphasize these estimates represent a heuristic argument based on several simplifying assumptions, they provide a plausible mechanism for the observed collision events at the main belt's inner edge. While this study focuses on the main belt as the source of the cs1 and cs2 events, we note that alternative scenarios can be responsible for the collision events. For example, a Solar System-like dynamical instability in Fomalhaut can cause planetesimals in the intermediate belt to be scattered and collide with bodies in the main belt, similar to \cite{Lawler2015}. We acknowledge that distinguishing between both scenarios would require detailed N-body, secular perturbation, and chaotic diffusion analyses to understand the mechanisms in which planetesimals can be perturbed out of the main belt into the orbits we describe. Such orbital modeling is beyond the scope of the present work and is left for future dynamical studies.

\subsection{Future Applications to Other Debris Disk Hosting Systems}

We envision the statistical model we have developed in order understand the collision rate of planetesimals in Fomalhaut being applied to a larger set of debris disks. It only requires the knowledge of basic disk properties like the belt location, width, eccentricity/relative velocity, and stellar mass. From these values, a collision rate for any planetesimal size can be extracted from a choice of planetesimal and belt properties. This can then be applied to specific disks that have been regularly imaged over time to place constraints on planetesimal belt properties and collision rates. Even non-detections of collisions in a particular disk can rule out certain bulk planetesimal properties. We encourage the regular monitoring of debris disks with HST in particular as it has archival data going back two decades and has demonstrated the capability to detect planetesimal collisions. We find compelling the most compelling sources to be dynamically active disks like $\beta$ Pictoris \citep[e.g.][]{Rebollido2024,Dent2014,Apai2015}, debris disks with ages similar to Fomalhaut, like HD 202628 and HD 207129 \citep[e.g.][]{Schneider2016}, and planetary systems with wide main belts ($\Delta a/a \ge 0.1$), like HD 181327 \citep[e.g.][]{Stark2014}, that allow for larger free eccentricities and, in turn, relative velocities where major collisions are likely to occur at younger ages than systems like Fomalhaut.

\subsection{Future Applications for Space-Based Exo-Earth Coronagraphic Imaging}

A key debate after the initial discovery of Fomalhaut cs1 (formerly Fomalhaut b) was whether the observed point source was due to a planet or other astrophysical phenomenon \citep[e.g.][]{Marengo2009}. It was only after multiple follow-up observations to pin down orbital properties and infrared emission in which it was identified as a collisional remnant \citep{Janson2012,Beust2014,Gaspar2020}. We see this as being an even more critical issue when it comes to Exo-Earth imaging with future missions like HWO. Specifically, the multiple order-of-magnitude contrast improvement expected from next generation space-based observatories compared to HST \citep[e.g.][]{Debes2019,Decadal2020} will be able to detect the collisional remnants from bodies that are multiple orders of magnitude more abundant (see Figure \ref{fig:pop}) than the planetesimal collisions seen by HST. Furthermore, the 5-10$\times$ increase in collision relative velocities \citep[e.g.][]{Bottke1994,Morbidelli2020} in inner planetary systems can sustain comparable collision rates to outer debris belts (i.e., exoKuiper Belts) even if the total planetesimal mass is smaller in the inner planetary system (see Equation \ref{totalrate}). This can translate to multiple point sources from planetesimal collisions appearing in a single Exo-Earth imaging epoch, which could require several follow-up observations to model the PSF and orbital motion of the point sources to distinguish between collisional remnants and an exoplanet.

For these reasons, the model we have developed in this study has utility in assisting future observations of Earth-like exoplanets with HWO. This model can be combined with constraints on the debris content (i.e., exozodi) of HWO targets to output the number of detectable planetesimal collisions that will be expected to masquerade as false positive Exo-Earth detections with HWO, which can then be used as an additional criteria to inform target selection and observing campaign designs.

\section{Conclusions}

In this study we have presented a technique to use planetesimal collision detections to set constraints on distribution of planetesimals and overall properties of debris belts. We then apply this to model to the Fomalhaut debris belt where two major collision events have been observed \citep{Kalas2008,Gaspar2020,Kalas2025}. The key results of this study are as follows:

\bigskip

(1) We present a simple statistical model that calculates the catastrophic collision rate of planetesimals as a function of six population parameters: the joint collisional cascade power-law index spanning both the strength and gravity dominated regime ($\alpha_{12}$), the power-law index governing the primordial planetesimal population ($\alpha_3$), the total mass of the belt ($M_{belt}$), the largest planetesimal present in the belt population ($s_{max}$), the break point where the planetesimal population transitions from the collisional evolved bodies to the primordial population ($s_{break}$), and the primordial population depletion parameter ($J$).

(2) We show that the monitoring for, and the discovery of collisional remnants in debris disks and belts has the potential to infer properties of the planetary system that cannot presently be directly detected or observed.

(3) By considering the two observed collisions by \cite{Kalas2008} and \cite{Kalas2025} (Fomalhaut cs1 and cs2), the $\sim$20-year observational baseline, the system age (440 Myr; \cite{Mamajek2012}), and the ALMA-derived dust mass (0.02 $M_{\oplus}$; \cite{Lovell2025}), we apply this model to the Fomalhaut system to constrain the belt parameters required to reproduce the observed events.

(4) The best fit belt and planetesimal properties assuming progenitor radii $\ge$ 100 km is as follows: a total belt mass of between 200-360 $M_{\oplus}$, the transition region between collisionally evolved and primordial planetesimal populations to be between 105-135 km in radius, and the largest planetesimal radius of 200-1000 km, with smaller radii favored over larger ones in our model.

(5) From the best fit belt and planetesimal properties, we derive a collision rate for target planetesimals with radii $\ge$ 100 km to be $0.086^{+0.067}_{-0.048}$ collision events per year in the region where cs1 and cs2 were observed to have originated.

(6) We find that for belt masses of 200-360 $M_{\oplus}$, we expect there to have been at least one collision, comparable to Fomalhaut cs1 and cs2, to have occurred in the observed Fomalhaut main belt in the past two decades. We explore the implications for a non-detection of this collision.

(7) We use our model to predict when we expect to find a Fomalhaut cs3, with a $\sim$50\% probability that the next comparable collision events occurs by $2031^{+10}_{-4}$ CE. Furthermore, we find Fomalhaut cs3 to be most likely to occur near the observed inner edge of the belt, near to discovery location of Fomalhaut cs2.

(8) We discuss how the continual monitoring of Fomalhaut and other debris disks with current and future observatories will test model predictions and inform future modeling.

(9) We discuss the utility of the developed collision model in future exo-Earth imaging, specifically to distinguish between planetary signals and potential false positives.

\section*{Acknowledgements}

We would like to thank Paul Kalas for discussions regarding the observations related to the discovery of Fomalhaut cs2. We also thank Namya Baijal for clarifying discussion regarding planetesimal impact physics. The results reported herein benefited from collaborations and/or information exchange within NASA's Nexus for Exoplanet System Science (NExSS) research coordination network sponsored by NASA's Science Mission Directorate. This material is based upon work supported by the National Aeronautics and Space Administration under Agreement No. 80NSSC21K0593 for the program ``Alien Earths''. Based on observations made with the NASA/ESA Hubble Space
Telescope, obtained at the Space Telescope Science Institute (STScI), which is operated by the
Association of Universities for Research in Astronomy, Inc., under NASA contract NAS 5-
26555.

\appendix

\section{Testing a Smoothed Surface Density Distribution in the Dynamically Hot Region Within the Observed Belt Inner Edge}\label{appendix-A}

An assumption in our model is that the surface density of planetesimals in the region where Fomalhaut cs1 and cs2 were found continues to follow the Gaussian surface density from the observed main belt. Given that the true distribution in this region is currently unknown, we test whether this assumption alters the best fit planetesimal population parameters, and the time/year when we expect the next collision to occur.

\begin{figure}
    \centering
    \includegraphics[width=0.5\linewidth]{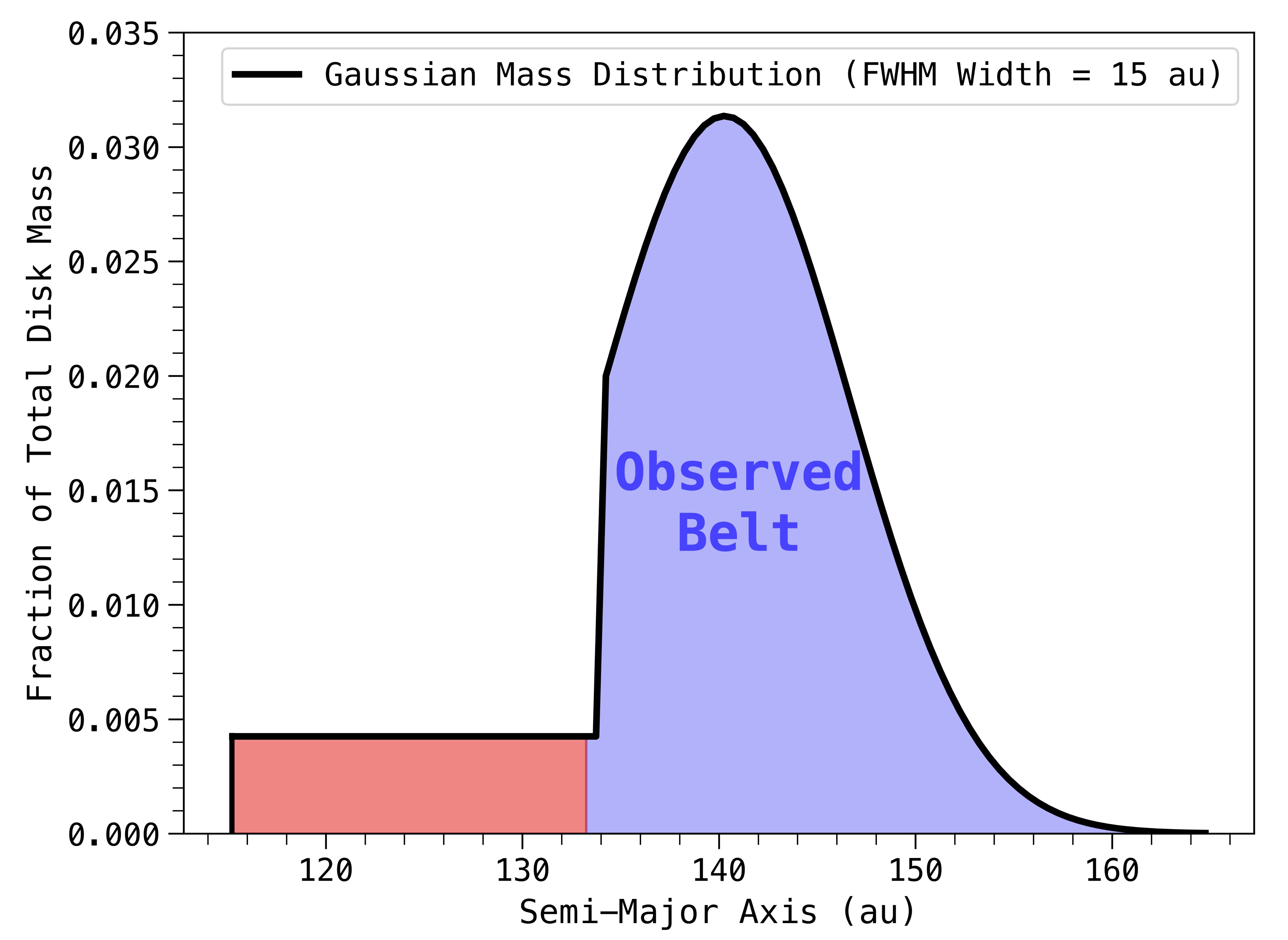}
    \caption{The fraction of the belt mass in each region of the disk as a function of belt semi-major axis, which is governed by a Gaussian surface density distribution in the dynamically cold, blue region and a smoothed/plateau surface density distribution in the dynamically hot, red region.}
    \label{fig:BeltSchematic-smoothed}
\end{figure}

To test this, we have modified our simulation with the surface density of planetesimals in the region within the belt inner edge to follow a smoothed/plateau surface density. The total surface density in this region is the same as the Gaussian version, but the surface density in the region is equally spread out (see Figure \ref{fig:BeltSchematic-smoothed}), rather than peaking at the observed belt inner edge. We find that the best fit planetesimal population parameters and year we expect the next collision to occur for both types of surface density distributions are consistent with one another within uncertainty. As more collisions are discovered in the Fomalhaut system, these will further constrain the shape of the surface density distribution in the region that is interior of the observed belt inner edge.

\begin{figure*}
    \centering
    \includegraphics[width=\linewidth]{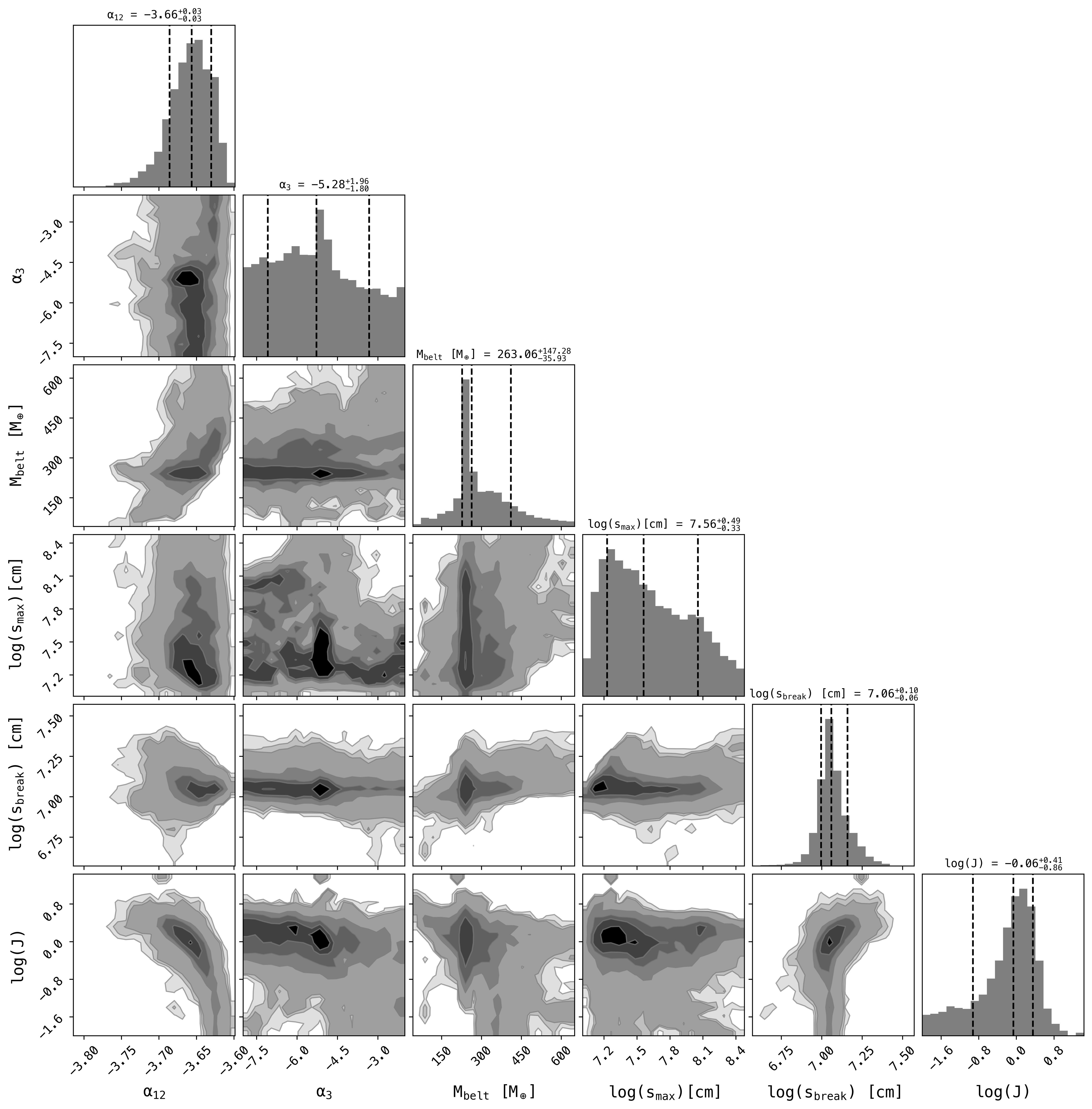}
    \caption{The posterior distribution of the {\tt\string emcee} chain of the six parameters fit in our model for collisional progenitors with radii $\ge$ 100 km and including a smoothed/plateau surface density. The model is fit with 50 walkers, 10000 steps, and 1000 burn-in steps. The dashed lines indicate the 16th, 50th, and 84th percentiles, respectively.  }
    \label{posterior100-smoothed}
\end{figure*}

\begin{figure*}
    \centering
    \includegraphics[width=0.9\linewidth]{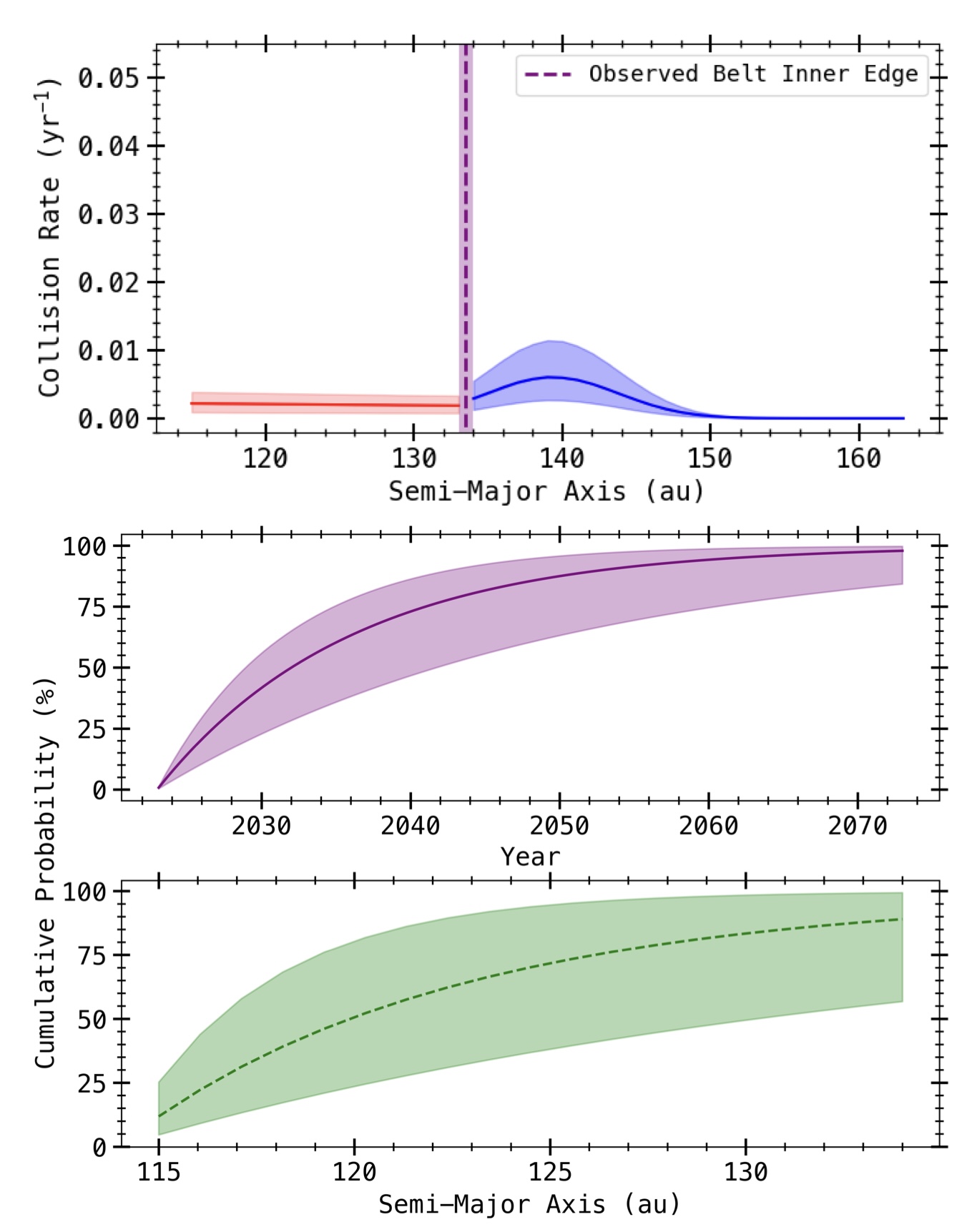}
    \caption{\textit{Top:} The catastrophic collision rate for target planetesimals $\ge$ 100 km in radius and a smoothed/plateau surface density distribution as a function of belt semi-major axis. The blue region indicates the observed Fomalhaut main belt, and the red region indicates the region where Fomalhaut cs1 and cs2 originated, which is within observed belt inner edge. \textit{Middle:} The cumulative probability of the next collision event (a hypothetical Fomalhaut cs3) where the target planetesimal radius is $\ge$ 100 km in radius and a smoothed/plateau surface density distribution as a function of collision year. The shaded regions indicate the 16th and 84th percentiles of the posterior distribution. We find a 50\% chance of the next collision occurring in 2027 at earliest and 2041 at the latest. \textit{Bottom:}  The cumulative probability of the next collision event (Fomalhaut cs3) where the target planetesimal radius is $\ge$ 100 km in radius and a smoothed/plateau surface density distribution as a function of belt semi-major axis.}
    \label{}
\end{figure*}

\bibliography{sample7}
\bibliographystyle{aasjournal}

\end{document}